\documentstyle[11pt]{article}

\begin{document}
%

\begin{center}
{\large \bf Using Dark Matter as a Guide to extend Standard Model:
Dirac Similarity Principle and Minimum Higgs Hypothesis}

\vskip.5cm

W-Y. Pauchy Hwang\footnote{Correspondence Author;
 Email: wyhwang@phys.ntu.edu.tw; arXiv:0106404 (hep-ph, 10 Sep 2010)} \\
{\em Asia Pacific Organization for Cosmology and Particle Astrophysics, \\
Institute of Astrophysics, Center for Theoretical Sciences,\\
and Department of Physics, National Taiwan University,
     Taipei 106, Taiwan}
\vskip.2cm


{\small(May 1, 2011)}
\end{center}

\begin{abstract}
We introduce the "Dirac similarity principle" that states that only those
point-like Dirac particles which can interact with the Dirac electron can
be observed, such as in the Standard Model. We emphasize that the existing
world of the Standard Model is a Dirac world satisfying the Dirac
similarity principle and believe that the immediate extension of the
Standard Model will remain to be so.
On the other hand, we are looking for Higgs particles for the last forty
years but something is yet to be found. This leads naturally to the
"minimum Higgs hypotheses".

Now we know firmly that neutrinos have tiny masses, but in the minimal
Standard Model there is no natural sources for such tiny masses. If nothing
else, this could be taken as the clue as the signature of the existence of
the extra heavy $Z^{\prime 0}$ since it requires the extra Higgs field, that
would help in generating the neutrino tiny masses.
Alternatively, we may have missed the right-hand sector for some reason. A simplified
version of the left-right symmetric electroweak model has the simplest Higgs sector
(in which there is one standard left-hand Higgs doublet together with the image
right-handed Higgs doublet), of which the Higgs sector also satisfies the "minimum
Higgs hypothesis". Or, if we question the origin of the family symmetry,
we may consider to enlarge it to the family gauge symmetry. It transpires very
feeble family interactions. All the above possibilities satisfy both the so-called
"Dirac similarity principle", the "minimum Higgs hypothesis", and renormalizability.

\bigskip

{\parindent=0pt PACS Indices: 12.60.-i (Models beyond the standard
model); 12.10.-g (Unified field theories and models); 14.70.Pw
(Other gauge bosons).}
\end{abstract}

\section{Dirac Similarity Principle}

Nowadays we all know that in the present Universe we have 25\% in dark matter
and only 5\% in "visible" ordinary matter, while living in the 70\% dark-energy
background. The Standard Model of particle physics so far describes only 5\% ordinary
matter. Let us review what we have: The Standard Model, that describes "point-like" Dirac
particles to begin with and so far ends up other "point-like" Dirac particles, summarizes
our experience of eighty years - "Dirac similarity principle". On the other hand, we have
looking very hard for the Higgs particle(s) for the last forty years - the experience may
be summarized as the "minimum Higgs hypothesis".

One way to describe our world is to follow the wisdom of Dirac - to describe
the electron as Dirac particle (spin-1/2). In fact, in Dirac's time, he wasn't
sure if he tried to describe a point-like relativistic particle. The logic which
he followed leads him a rather unique equation - the Dirac equation\cite{Wu}. To be
"strange", he combined the orbital momentum (in terms of the space-time notion)
with the Pauli matrix (a pure number) to explain half spins (and quantum mechanics).
To describe the world in accord with relativity, we would introduce the antiparticle
of the electron - the positron.

Now, what are the others? Then maybe the neutron and proton, but they are
extended, so instead we think of quarks, the constituents of hadrons and it turns
out to be also the so-called "point-like Dirac particles". So, we come to the
Standard Model where quarks and leptons are of three generations; all are
point-like Dirac particles but described by quantized Dirac fields. Furthermore,
they interact via quantized gauge fields (spin-1), and so far we haven't observed
any scalar (Higgs) fields (spin-0).

The world of the Standard Model, our world, seems to be extremely simple.
One starts out with the electron, a point-like spin-1/2 particle, and ends
up with other point-like Dirac particles, but with interactions through
gauge fields modulated by the Higgs fields. This is what we experimentally
know, and it is a little strange that it seems to be "complete" and that
nothing else seems to exist. After Dirac's equation and search for the
point-like particles, now for eighty years, Dirac equations explain all the
relativistic point-like particles and their interactions.

The "Dirac similarity principle" states that every "point-like"
or quantized particle of spin-1/2 could be observed if it is "connected"
with the electron, the original spin-1/2 particle. For some reason\cite{Wu}, this
has something to do with how relativity and the space-time structure gets married
with spin-1/2 particles. This is interesting since there are other ways to write
spin-1/2 particles, but so far they are not seen perhaps because they are not
connected with the electron.

These are "point-like" Dirac particles of which the size we believe is less than
$10^{-17}\,cm$, the current resolution of the length. Mathematically, the
"point-like" Dirac particles are described by "quantized Dirac fields" - maybe
via renormalizable lagrangian. The "quantized Dirac fields", which we can
axiomatize for its meaning, in fact does not contain anything characterizing
the size (maybe as it should be) - maybe the true meaning of "point-like Dirac
particles" ought to be. Quantum mechanics makes "point-like particles" as
they are supposed to be. Of course, the word "renormalization" contains something of
the variable size.

\section{Minimum Higgs Hypothesis}

Let's now "apply" the Dirac similarity principle. The neutrinos
are now Dirac particles of some kind - so, right-handed neutrinos
exist and the masses could be written in terms of them. To make
Dirac neutrinos massive, we need a Higgs doublet for that. Is this
Higgs doublet a new Higgs doublet? In principle, we could use the
Standard-Model Higgs doublet and take and use the complex conjugate
(like in the case of quark masses) - the problem is the tininess of
these masses and if this would go it is definitely un-natural.

Thus, we would assume that a new and "remote" Higgs doublet would exist and
that the tininess of the neutrino masses is explained by the neutrino
couplings to the "remote" Higgs. Then it is "natural" to get these tiny masses.
Why are the neutrino couplings to "remote" Higgs doublet
should be small? - just similar to the CKM matrix elements (that is, the 31
matrix element is much small than the 21 matrix element); the other
"naturalness" reason.

Thus, we may introduce the "minimum Higgs hypothesis" as anther working
conjecture. We should look for the minimum number of Higgs multiplets
as our choice and the couplings to the "remote" Higgs would be much smaller
than to the ordinary one. As said earlier, this hypothesis makes the case of
the tiny neutrino masses very natural and, vice versa, we rephrase the natural
situation to get the hypothesis. Why do we adopt such hypothesis? For more than
forty years, we haven't found the solid signature for the Higgs; that the neutrinos
have tiny masses (by comparison with quarks and charged leptons) is another reason.

{\bf Therefore, under "Dirac similarity principle" and "minimum Higgs hypothesis", we have
a unique Standard Model if the gauge group is determined or given.}

That neutrinos have tiny masses can be taken as a signature that there is a heavy
extra $Z^{\prime 0}$, so that a new Higgs doublet should exist. This extra
$Z^{\prime 0}$ then requires the new "remote" Higgs doublet\cite{Hwang}.
This Higgs doublet also generates the tiny neutrino masses. On other hand,
we could require that the right-hand $SU_R(2)$ group, which contains the extra
$Z^{\prime 0}$, exists to restore the left-right symmetry. In all cases, we
need the extra $Z^{\prime 0}$ for the sake that there exists a new "remote"
Higgs doublet that makes neutrino masses in a natural way. But we should follow
the "minimum Higgs hypothesis" to have one standard Higgs doublet and another
new "remote" Higgs doublet, to complete the story.
We note that they all may belong to the so-called "dark matter", 25\%
of the present-day Universe (compared to 5\% of ordinary matter).

In a world of point-like Dirac particles interacting with the generalized
standard-model interactions, there are left-handed neutrinos belong to
$SU_L(2)$ doublets while the right-handed neutrinos are singlets.
The term specified by
\begin{equation}
\varepsilon\cdot ({\bar\nu}_L,{\bar e}^-) \nu_R \varphi
\end{equation}
with $\varphi=(\varphi^0,\varphi^-)$ the new Higgs doublet could generate
the tiny mass for the neutrino.

As said earlier, we introduce the "minimum Higgs hypothesis". To the first
(standard) Higgs doublet, from the electron to the top quark we call it
"normal" and $G_i$ is the coupling to the first Higgs doublet, and to the
second (extra, or "remote") Higgs doublet the strength of the couplings for the Dirac
particles is down by the factor $(v/v')^2$ with $v$ the VEV for the standard Higgs
and $v'$ the VEV for the remote Higgs. The hypothesis sounds very reasonable, similar
to the CKM matrix elements, and one may argue about the second power but for the
second Higgs fields some sort of scaling may apply.

With the working hypothesis, the coupling of the neutrinos to the standard
Higgs would vanish completely (i.e., it is natural) and its coupling to the
second (remote) Higgs would be $G_j (v/v')^2$ with $G_j$ the "normal" size.

The "minimum Higgs hypothesis" amounts to the assertion that there should be as
least Higgs fields as possible and the couplings would be ordering like the above
equation, Eq. (1). This conjecture arises from the fact that over forty years we
are still looking for the first signature of Higgs particle(s).

Indeed, in the real world, neutrino masses are tiny
with the heaviest in the order of $0.1\, eV$. The electron, the lightest
Dirac particle except neutrinos, is $0.511\, MeV$\cite{PDG} or $5.11 \times 10^5\, eV$.
That is why the standard-model Higgs, which "explains" the masses of all other
Dirac particles, is likely not responsible for the tiny masses of the neutrinos.
The "minimum Higgs hypothesis" sort of makes the hierarchy very natural.

In an early paper in 1987\cite{Hwang}, we studied the extra $Z^{\prime 0}$
extension paying specific attention to the Higgs sector - since in the Minimal
Standard model the standard Higgs doublet $\Phi$ has been used up by $(W^\pm,\,Z^0)$.
We worked out by adding one Higgs singlet (in the so-called 2+1 Higgs scenario) or
adding a Higgs doublet (the 2+2 Higgs scenario). It is the latter that we could add the
neutrino mass term naturally. (See Ref.\cite{Hwang} for details. Note that the
complex conjugate of the second "remote" Higgs doublet there is the $\varphi$ above.)

The new Higgs potential follows the standard Higgs potential, except
that the parameters are chosen such that the masses of the new Higgs are much
bigger. The coupling between the two Higgs doublets should not be too big to
upset the nice fitting\cite{PDG} of the data to the Standard Model. All
these go with the smallness of the neutrino masses. Note that spontaneous symmetry
breaking happens such that the three components of the standard Higgs get absorbed
as the longitudinal components of the standard $W^\pm$ and $Z^0$.

In this junction, we should say something about the cancelation of the
flavor-changing scalar neutral quark currents. Suppose that we work with two
generations of quarks, and it
is trivial to generalize to the physical case of three. I would write
\begin{eqnarray}
({\bar u}_L,\,{\bar d}^\prime_L)d^\prime_R\Phi + c.c.;\nonumber\\
({\bar c}_L,\,{\bar s}^\prime_L)s^\prime_R\Phi + c.c.;\nonumber\\
({\bar u}_L,\,{\bar d}^\prime_L)u_R \Phi^*+c.c.;\nonumber\\
({\bar c}_L,\,{\bar s}^\prime_L)c_R \Phi^*+c.c.,
\end{eqnarray}
noting that we use the rotated down quarks and we also use the complex conjugate
of the standard Higgs doublet. This is a way to ensure that
the GIM mechanism\cite{GIM} is
complete. Without anything to do the opposite, I think that it is reasonable to
continue to assume the GIM mechanism.

Coming back to think about it seriously, the standard Higgs doublet doing the
play in the quark sector is another play of the "minimum Higgs hypoyhesis".
Otherwise, there could be many more Higgs doing the plays for us.

\section{A World of "Point-like" Dirac Particles}

We may rephrase the Standard Model as a world of
point-like Dirac particles, or quantized Dirac fields, with interactions.
Dirac, in his relativistic construction of Dirac equations, was enormously
successful in describing the electron. (The point-like nature of the electron was
realized almost a century later.) Quarks, carrying other intrinsic degrees (color),
are described by Dirac equations and interact with the electron via gauge fields.
We also know muons and tau-ons, the other charged leptons. So, how about neutrinos?
Our first guess is also that neutrinos are point-like Dirac particles of some sort
(against Majorana or other Weyl fields). For some reasons, point-like Dirac
particles in our space-time are implemented with certain properties - that they
know the other point-like Dirac particles in our space-time. That is why we call it
the "Dirac similarity principle" to begin with. This is a world of point-like Dirac
particles interacting among themselves via gauge fields modulated by Higgs fields.

To sum up, for our real world, we begin with the electron and end up with all
three family of quarks and leptons, with gauge fields of strong and electroweak
interactions modulated by Higgs fields - the world satisfied by the
Dirac similarity principle. To proceed from there, we treat neutrinos
as Dirac particles in accord with the Dirac similarity principle.

As a matter of fact, we will treat only with renormalizable interactions
- with the spin-1/2 field power 3/2 and scalar and gauge fields power 1;
the total power counting less than 4. What is surprising is the role of
"renormalizability". We could construct quite a few such extensions of
the minimal Standard Model; they are all renormalizable - the present
extra $Z^{\prime 0}$\cite{Hwang}, the left-right model (in the minimum 
sense), and the recent proposed family gauge 
theory\cite{Family}; there are more. Apparently, we should not give up 
though the road seems to have been blocked. 

Our space-time is "defined" when the so-called "point-like Dirac particles"
are "defined", and vice versa. On the other hand, "point-like Dirac particles"
are in terms of "quantized Dirac fields". These concepts are "defined" together,
rather consistently.

\section{A Simple Connection to the Unknowns}

As emphasized earlier, a world of point-like Dirac particles, as
described by quantum field theory (the mathematical language), turns out
to be the physical world around us - that may also define the space-time for us.
The interactions are mediated by gauge fields modulated slightly by Higgs
fields. There may be some new gauge fields, such as the extra $Z^{\prime 0}$,
or the missing right-handed partners\cite{Salam}, or the family gauge
symmetry\cite{Family}, or others. The first important experimental clue may be
the tiny neutrino masses, suggesting that there is at least\cite{Hwang} one extra
$Z^{\prime 0}$. To proceed further, the missing right-hand sector\cite{Salam}
might come back. If certain symmetry cannot find any reasons, such as the family
symmetry, it can be gauged\cite{Family} - thus connecting with very feeble
interactions.

There are a bunch of dark matter out there - 25\% of the present Universe
compared to only 5\% ordinary matter. The new gauge bosons, together with Higgs
particles and others, can be identified as "dark-matter particles", due to the
proposed feeble interactions. When we get more knowledge on the dark
matter, we may have pretty good handles of this Universe. Of course, neutrinos can
be viewed also as one kind of dark matter (both from their feeble interactions
with ordinary matter species and from whether the family symmetry could be gauged)
- so, hopefully from neutrinos we can peel into a bigger world of dark matter.

There are two related remarks. The first remark related to the $SU_L(2)\times
SU_R(2) \times U(1)$ model\cite{Salam}. The second has to be related to the
family (gauge) symmetry\cite{Family}.

In the $SU_L(2)\times SU_R(2) \times U(1)$ model\cite{Salam}, suppose that in the
left and right parts each has one Higgs doublet (minimal) and we could try
to make the tiny neutrino masses in the right-handed sector. Here we employ the
"minimum" working hypothesis. This seems to be
rather natural. We should think about this possibility very seriously - except
that we should think of the Higgs mechanism in a real minimum fashion, judging
that we are looking for Higgs for about forty years. Thus, we have one left-handed
Higgs doublet and another right-handed (remote) Higgs doublet - due to spontaneous
symmetry breaking (SSB), only two neutral Higgs particles are left. That means that
we are advocating a particular kind of the left-right symmetric model.

Regarding the family (gauge) symmetry, it is difficult to think
of the underlying reasons why there are three generations (of quarks and leptons).
This is why we promote the family symmetry to the family gauge
symmetry\cite{Family}. In both cases, the proposed Dirac similarity principle
and the "minimum Higgs hypothesis" both
may hold - an interesting and strange fact.

In both cases in the above remarks, it implies that, at temperature somewhat
higher than $1\,TeV$, there would be another phase transition - for the
spontaneous symmetry breaking. If most of the products from the phase transition
would remain to be dark matter, we would have most natural candidates for
the dark matter.

In other words, those unseen particles, owing to their weak interactions with
ordinary matter, can be classified as "dark matter" in the extra $Z^{\prime 0}$
model, or in the left-right model, or in the family gauge symmetry model.

The other interesting aspect is that the left-right symmetry is the missing
symmetry while the family gauge symmetry is the symmetry which we have found
but suspect that it is not complete.

\section*{Acknowledgments}
This research is supported in part by National Science Council project (NSC
99-2112-M-002-009-MY3).

\end{document}